\documentclass[journal=jpclcd,manuscript=perspective]{achemso}

\usepackage[T1]{fontenc} 
\usepackage[version=3]{mhchem}
\usepackage{xcolor}
\usepackage{paralist} 

\author{Ajay~Ram~Srimath~Kandada}
\affiliation[IIT]
{Center for Nano Science and Technology@PoliMi, Istituto Italiano di Tecnologia, via Giovanni Pascoli 70/3, 20133 Milano, Italy}
\alsoaffiliation[WFU]
{Department of Physics, Wake Forest University, P.O.\ Box 7507, Winston-Salem, NC~27109, United~States}
\email{srimatar@wfu.edu}

\author{Carlos~Silva}
\affiliation[GAtech-chem]
{School of Chemistry and Biochemistry, Georgia Institute of Technology, 901 Atlantic Drive, Atlanta, GA~30332, United States}
\alsoaffiliation[GAtech-phys]
{School of Physics, Georgia Institute of Technology, 837 State Street NW, Atlanta, GA~30332, United States}
\alsoaffiliation[GT-MSE]
{School of Materials Science and Engineering, Georgia Institute of Technology, North Avenue, Atlanta, GA 30332, United~States}
\email{carlos.silva@gatech.edu}

\title[Exciton polarons in 2D hybrid perovskites]
  {Perspective: Exciton polarons in two-dimensional hybrid metal-halide perovskites}

\begin{document}

\begin{tocentry}
%
%
%
\begin{center}
    \includegraphics[width=5cm]{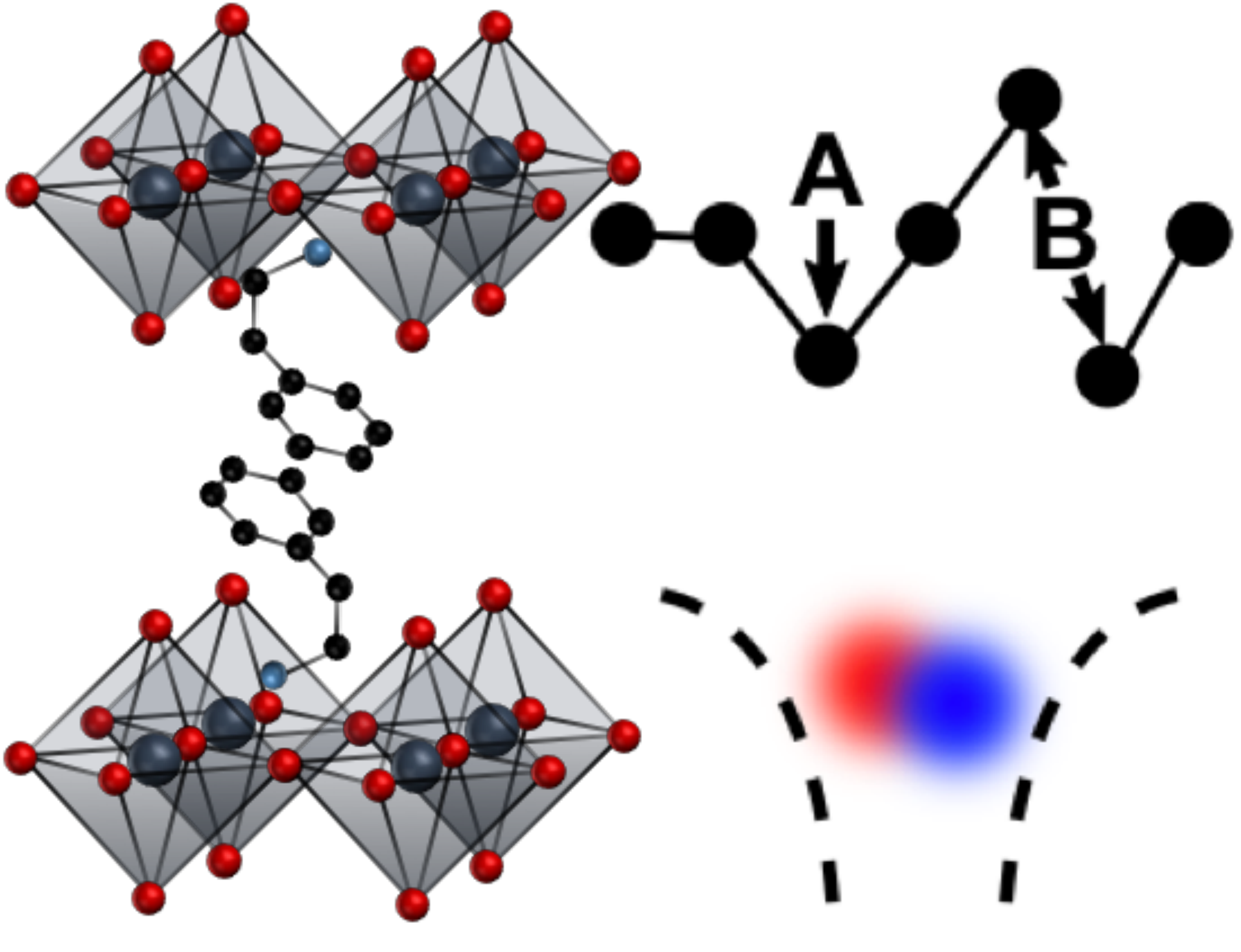}
\end{center}
Exciton polarons in 2D hybrid organic-inorganic metal-halide perovskites.
\end{tocentry}

\begin{abstract}
While polarons --- charges bound to a lattice deformation induced by electron-phonon coupling --- are primary photoexcitations at room temperature in bulk metal-halide hybrid organic-inorganic perovskites (HOIP), excitons --- Coulomb-bound el\-ectron-hole pairs --- are the stable quasi-particles in their two-dimensional (2D) analogues. Here we address the fundamental question: are polaronic effects consequential for excitons in 2D-HIOPs? Based on our recent work, we argue that polaronic effects are manifested intrinsically in the exciton spectral structure, which is comprised of multiple non-degenerate resonances with constant inter-peak energy spacing. We highlight our own measurements of population and dephasing dynamics that point to the apparently deterministic role of polaronic effects in excitonic properties. We contend that an interplay of long-range and short-range exciton-lattice couplings give rise to exciton polarons,  a character that fundamentally establishes their effective mass and radius, and consequently, their quantum dynamics. Finally, we highlight opportunities for the community to develop  the rigorous description of exciton polarons in 2D-HIOPs to advance their fundamental understanding as model systems for condensed-phase materials in which lattice-mediated correlations are fundamental to their physical properties.
\end{abstract}


\section{Preamble}

Metal-halide hybrid organic-inorganic perovskites (HOIPs) are direct bandgap semiconductors with rich, complex photophysics compared to established semiconductors such as III-V composite materials~\cite{stranks2015metal, srimath2016photophysics, egger2018remains}. A deterministic element of these physics stems from the strong electron-phonon coupling arising from their ionic character, and also from the convoluted dynamics of the hybrid organic-inorganic lattice, which is soft and highly noisy~\cite{bakulin2015real,zhu2016screening, zhu2017organic}. These effects have a substantial influence on electronic excitations and are at the core of investigations of photoexcitation dynamics in HOIPs. Key to this understanding is that polarons --- charges dressed by specific phonons and bound to the lattice deformation induced by Coulomb forces --- are the primary excitations in HOIPs. Polaronic effects have been suggested to play an important role in the excitation dyanamics and carrier transport in lead-halide HOIPs: the relatively long carrier lifetimes with small bimolecular recombination coefficient~\cite{wehrenfennig2014high}, slow thermalization dynamics~\cite{niesner2016persistent, zhu2017organic} of hot carriers and weak dephasing rates~\cite{march2017four} are believed to originate from the protection offered to the charged excitations by the dressing of the lattice phonons~\cite{miyata2017large}. The initial hypothesis of charge carriers as polarons in HOIPs was proposed based on transport measurements, which revealed modest carrier mobilities~\cite{yi2016intrinsic}. Further experimental and theoretical verifications subsequently emerged. Vibrational spectra obtained via time-domain experiments involving impulsive excitations revealed a distinct lattice configuration in the presence of photo-excited species~\cite{batignani2018probing}. Time-resolved optical Kerr effects suggested that polaron formation time is within a picosecond, with a non-trivial dependence on the nature of the structure of the coordinating cation~\cite{zhu2016screening}. Correlated motion of charge excitations with the lattice motion is also demonstrated via optical pump, THz probe spectroscopy~\cite{lan2018ultrafast}. The uniqueness of the carrier-lattice coupling is further attributed to the role of dielectric relaxation effects from the dynamic fluctuations and lattice anharmonic effects~\cite{guo2019dynamic, yaffe2017local, bonn2017role}. 
These peculiar lattice interactions are thus considered to play a primary role in protecting the photo-excitations from non-linear scattering and recombination processes, driven by a dynamic screening of the Coulomb interactions between carriers by the ionic lattice fluctuations. A natural corollary to such a scenario is the screening of electron-hole binding interactions~\cite{even2014analysis}, and, in fact, the reported exciton binding energies are in the order of the lattice thermal energy $k_B T$ at ambient conditions in bulk HOIPs~\cite{miyata2015direct}, resulting in unstable excitons at room temperature. 

The effects of such an electronic-vibrational landscape undergo substantial transformation in two-dimensional (2D) HOIP derivatives, which are multiple-quantum-well-like derivatives of HOIPs~\cite{saparov2016organic}. These are composed of quasi-2D layers of metal-halide lattice planes that are separated by long organic cationic spacers with an average inter-layer separation of $\sim 1$\,nm. Due to negligible contributions to the frontier orbitals from the organic cation and to the absence of orbital overlap between the metal-halide layers, electronic excitations are confined within the 2D inorganic sub-lattice~\cite{even2014understanding, gauthron2010optical, diab2016narrow}. Due to the large dielectric contrast between the organic and inorganic layers, Coulomb correlations are enhanced dramatically as a result of image-charge effects, and the dynamic Coulomb screening effects are also mainly confined to the 2D lattice plane defined by the inorganic sub-lattice. This results in exciton binding energies $\sim 10\,k_B T$ at room temperature~\cite{blancon2018scaling, straus2016direct, Straus2018a, pedesseau2014electronic}. In this perspective, we focus on the consequences of the fluctuating lattice interactions on the excitonic characterstics, focusing on their linear and non-linear spectral structures and on population and dephasing dynamics. In particular, we argue that polaronic effects are not only active, but that they fundamentally define excitons in 2D-HIOPs. We will argue that the phonon coupling associated with polarons are in a unique intermediate regime between that of covalent crystals and molecular systems, giving rise to exciton polarons, which makes the formal description of these quasi-particles challenging, but which also makes such description of high fundamental importance for the development of the semiconductor understanding of hybrid metal-halide semiconductors. Our perspective is that the research highlighted in this article presents a substantial challenge for the electronic structure and quantum dynamics communities interested in the properties of hybrid, ionic semiconductor materials.

\section{Exciton spectral structure}

\begin{figure}[ht]
\centering
\includegraphics[width=8.25cm]{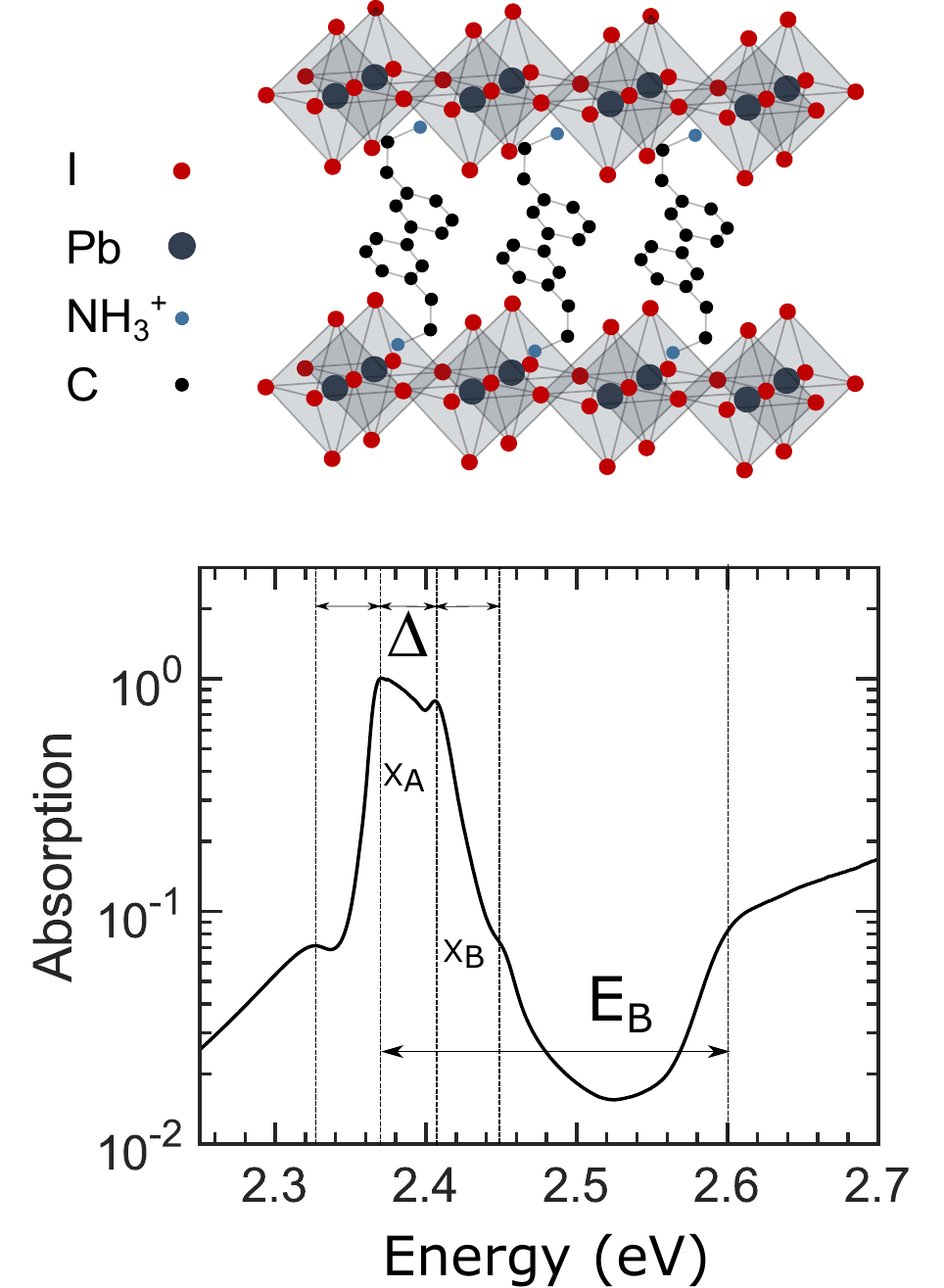}
\caption{Crystal structure of a prototypical 2D perovskite: phenylethylammonium lead iodide (\ce{(PEA)2PbI4}). (Bottom) Linear absorption spectrum of \ce{(PEA)2PbI4} taken at $T= 5$\,K; $\Delta \sim 35\pm 5$\,meV represents the energy spacing within the excitonic finestructure and $E_B \sim 250$\,meV is the exciton binding energy associated with the main exciton peak. }
\label{fig:linear}
\end{figure}

The context of our perspective lies in the complex lineshape of 2D-HIOPs. The crystal structure of \ce{(PEA)2PbI4} (PEA = phenylethylammonium), a prototypical 2D-HOIP, is shown in Fig.~\ref{fig:linear}. Exciton resonances can be observed in the optical absorption spectra well below the continuum edge, see Fig.~\ref{fig:linear}. The excitonic lineshape is composed of a peculiar finestructure with at least four distinguishable peaks that are equally separated by $\Delta \approx 35$--40\,meV, as shown in Fig.~\ref{fig:linear}. Such a spectral finestructure within the excitonic band has been reported for multiple 2D-HOIPs containing a variety of organic cations and halogens~\cite{Ishihara1990, Ishihara1992, Goto2006, Kitazawa2010, Kitazawa2010a, Kitazawa2011, Kitazawa2012, Kataoka1993, Shimizu2005,Shimizu2006} with various plausible explanations of its origin~\cite{kataoka1993magneto, Ema2006, Straus2018a, takagi2013influence,mauck2019excitons}, yet with no clear consensus. We have recently addressed the nature and origin of these distinct excitonic states over a series of publications (Refs.~\citenum{Thouin2018,Neutzner2018,thouin2019phonon,thouin2019polaron,thouin2019enhanced}) based on various linear and non-linear optical spectroscopies, where we have identified an unequivocal correlation with lattice interactions that led us to the hypothesis of exciton polarons as the primary photoexcitation in 2D-HOIPs. The starting point of our discussion is that we rule out a vibronic progression of a single exciton as the origin of this finestructure, and we instead claim that it arises from a family of co-existing, correlated excitons with distinct binding energy that are intrinsic to the electronic structure. The principal phenomenology stemming from our work that has led us to this view is the following:  
\begin{inparaenum}[(i)]
	\item we established, by means of coherent two-dimensional excitation spectroscopy, that the exciton spectral structure in Fig.~\ref{fig:linear} reflects multiple correlated transitions involving a common ground state~\cite{Thouin2018};
	\item we also observed exciton coherences via transitions to biexciton states that oscillate in population-waiting time with period $\hbar/\Delta$~\cite{Neutzner2018}, reinforcing that $\Delta$ is intrinsic to the excitonic structure;  
	\item we measured, by means of resonant impulsive stimulated Raman scattering, the phonon spectrum associated with motion in the metal-halide sub-lattice that couples to the multiple excitons distinctly~\cite{thouin2019phonon}, and found that these dressing phonons play an active role in inter-exciton nonadiabatic conversion~\cite{thouin2019polaron}; 
	\item we quantified contrasting biexciton interactions for particular excitons within the finestructure, with the principal exciton showing weakly repulsive correlations, while its higher-energy counterpart displays a binding energy of $\sim 45$\,meV at room temperature~\cite{Thouin2018}; 
	\item we evaluated different multi-exciton Coulomb-mediated elastic scattering rates for distinct excitons within the spectral structure, with these dynamics activated by different phonons for different excitons~\cite{thouin2019enhanced}. 
\end{inparaenum}
Therefore, \emph{while the spectral structure in Fig.~\ref{fig:linear} is intrinsic, the dynamical and multi-particle interaction properties of different excitons probed via distinct resonances within the finestructure, are unique, establishing that the spectral structure portrays a family of co-existing excitons with bidning energy offset by $\Delta$, each with distinct lattice dressing}. We will outline these observations in what follows, and then we will turn to a discussion of polaronic effects by reviewing established formalisms for the description of polarons and exciton polarons in order to invoke these concepts in the rationalization of our reported phenomenology.

\subsection{Exciton coherent spectral signatures and dynamics}

\begin{figure}[ht]
\centering
\includegraphics[width =10.5cm]{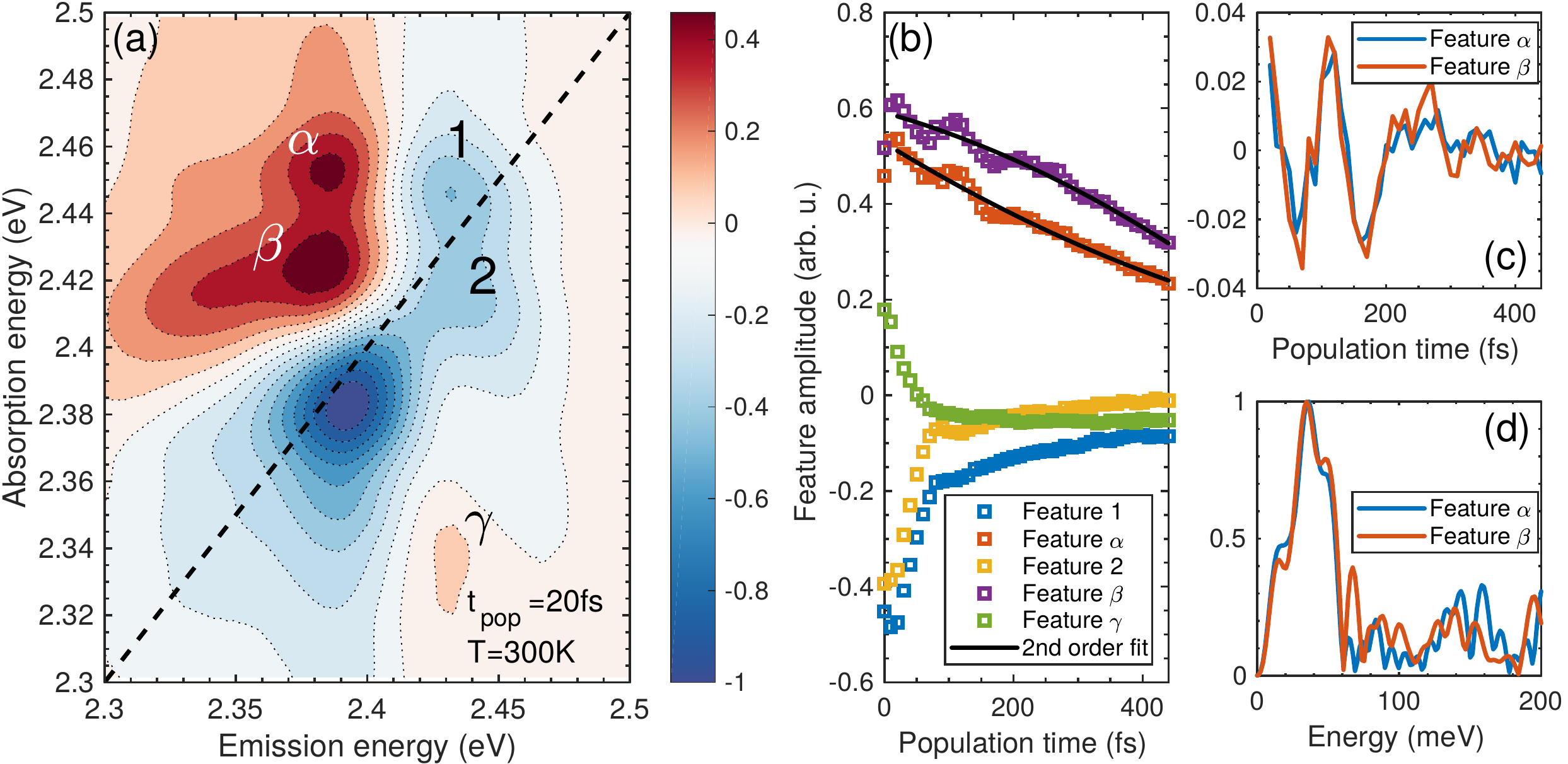}
\caption{(a) Total correlation 2D coherent excitation spectrum taken at room temperature for a population time delay of 20\,fs. (b) Signal of various features labelled in (a) as a function of population time delay. The $\alpha$ and $\beta$ features are fitted with a quadratic polynomial to isolate the oscillatory components. The residual of this fit is shown in (c) and the norm of its Fourier transform is shown in (d). Figure reproduced with permission from ref.~\citenum{Neutzner2018}}
\label{fig:dynamics}
\end{figure}

We begin by considering the coherent two-dimensional excitation lineshape of a \ce{(PEA)2PbI4} polycrystalline film in  Fig.~\ref{fig:dynamics}, reproduced from Ref.~\citenum{Neutzner2018}. An extensive assignment of the observed features in Fig.~\ref{fig:dynamics}(a) can be found in Refs.~\citenum{Thouin2018} and \citenum{Neutzner2018}; here we highlight the salient points for the purpose of laying out our perspective. We identify off-diagonal cross peaks between exciton diagonal peaks, indicating that the various excitons share a common ground state. In fact, in Ref.~\citenum{thouin2019enhanced}, we present data measured at substantially lower fluence than that extracted from Ref.~\citenum{Neutzner2018} in Fig.~\ref{fig:dynamics}, and the 2D coherent excitation lineshape, including the cross-peak structure, is more clear at lower exciton density, in which exciton-exciton collisional contrbutions to line broadening are limited (discussed below)~\cite{thouin2019enhanced}. Secondly, the oscillatory dynamics of off-diagonal features $\alpha$ and $\beta$, assigned to biexciton coherences~\cite{Thouin2018}, are evident in Fig.~\ref{fig:linear}(b). A Fourier analysis of the dynamics revealed an energy of 35\,meV, the same as the inter-excitonic spacing $\Delta$, and thus interpreted as the signature of inter-excitonic coherence. Such an observation is an unambiguous demonstration for the existence of not only a common ground state but also of common higher-lying states. 

\subsection{Strong exciton-lattice coupling}

\begin{figure}[ht]
	\centering
	\includegraphics[width=\textwidth]{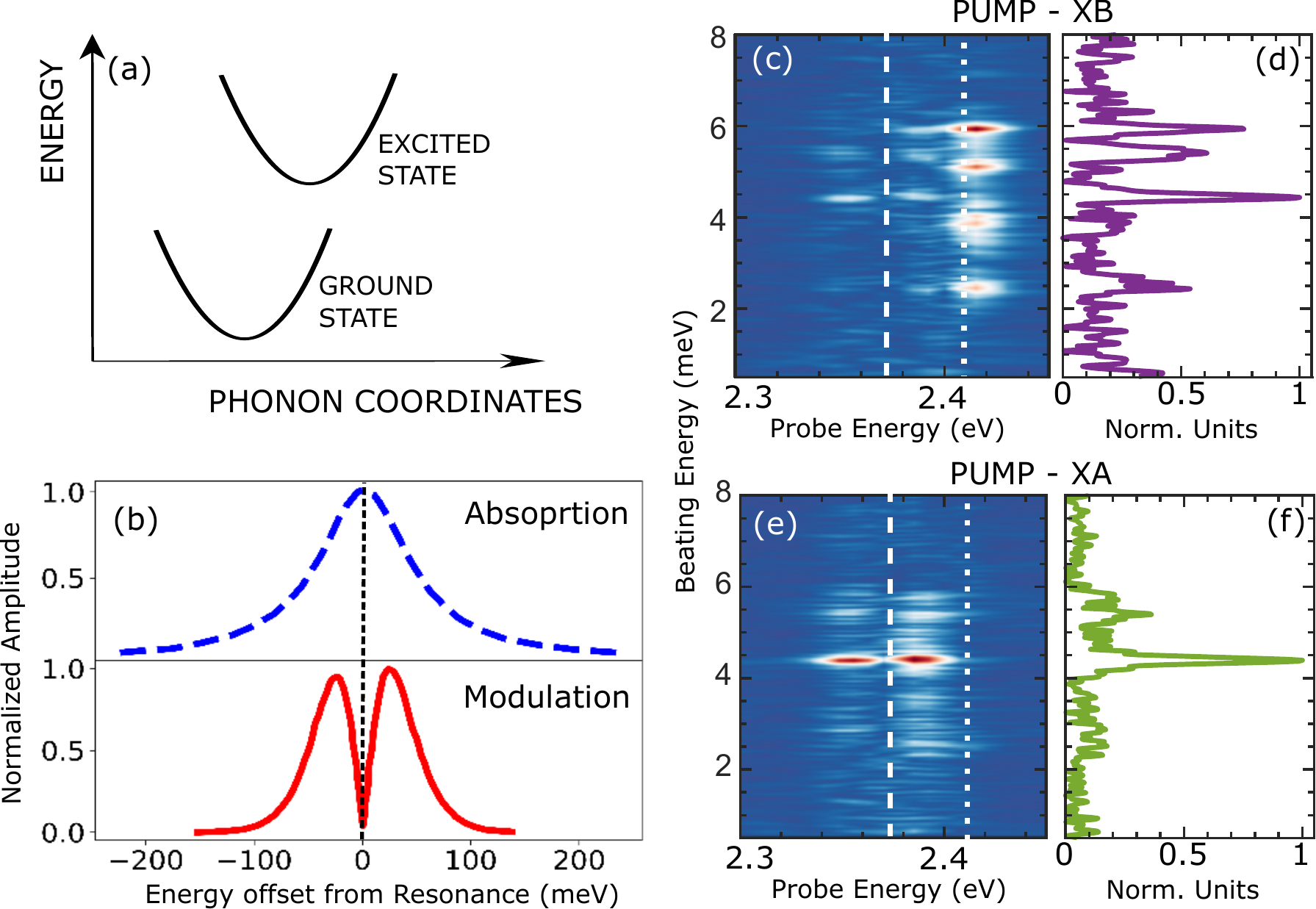}
	\caption{(a) Schematic of potential energy surfaces for exciton polarons along generic phonon direct-space coordinates. (b) Schematic of the amplitude spectrum of phonon coherences induced by resonant impulsive stimulated Raman scattering (RISRS) at the excitonic resonance. (c) and (d) RISRS energy versus probe photon energy map and probe-energy-integrated RISRS amplitude versus phonon energy, respectively, upon resonant pumping of $X_B$. (e) and (f) RISRS energy versus probe photon energy map and probe-energy-integrated RISRS amplitude versus phonon energy, respectively, upon resonant pumping of $X_A$. Data extracted from Ref.~\citenum{thouin2019phonon}.}
	\label{fig_natmater}
\end{figure}

Our analysis of the linear and nonlinear optical lineshapes of \ce{(PEA)2PbI4}, prompted by the apparent independence of $\Delta$ with the number of lead iodide lattice layers, and on the identity of the organic cation~\cite{Neutzner2018}, and by the vicinity of $\Delta$ to polaron binding energies~\cite{munson2019lattice,mahata2019large}, led us to hypothesize that polaronic effects could contribute to the exciton spectral finestructure. In order to probe exciton-phonon coupling details, we employed resonant impulsive stimulated Raman spectroscopy, which involves impulsive optical pumping of coherent lattice motion coupled to the electronic excitation. Such process produces a coherent vibrational wavepacket (a coherent sum of all the vibrational modes of the lattice) that oscillates within the potential energy surfaces of the ground and excited states along the coordinates defined by various Raman-active phonon modes, see Fig.~\ref{fig_natmater}(a). Subsequently, optical absorption around the exciton resonance is modulated at the frequency of the lattice motion, which can be perceived by the transmission of a time-delayed probe pulse. The Fourier transform of such time-oscillating dynamics produces the impulsive stimulated Raman spectrum. As shown in Fig.~\ref{fig_natmater}(b), the amplitude spectrum of the modulation along the probe energy exhibits a characteristic dip at the peak energy of the exciton resonance and is a direct consequence of the coupling of the impulsively excited lattice modes with that particular optical transition. 

We have performed such experiments on prototypical 2D-HIOPs and the details are reported in Ref.~\citenum{thouin2019phonon}. The most surprising aspect of our work lies not in the observation of coherent phonons upon resonant excitation of photocarriers and excitons, but in the distinctness of the dynamics when exciting different excitons within the finestructure. Let us consider the two dominant excitonic lines in Fig.~\ref{fig:linear}, which we label $X_A$  and $X_B$,  peaked at 2.36 and 2.41\,eV, respectively. Fig.~\ref{fig_natmater}(c) shows beating maps plotted as a function of the probe energy and the phonon energy when $X_B$ is excited, and Fig.~\ref{fig_natmater}(d) is that when $X_A$ is resonantly pumped. Figs.~\ref{fig_natmater}(d) and (f) are spectrally integrated over probe photon energy. The differences in their respective behavior is striking. The coherent modes excited in both cases are vastly different and their motion either modulates $X_A$ or $X_B$ exclusively. These observations led us to two important inferences: 
\begin{inparaenum}[(1)]
	\item a displaced oscillator model as sketched in Fig.~\ref{fig_natmater}(a) can be invoked in these systems despite the softness and substantial anharmonicity of the lattice, and 
	\item there are energetically close and correlated (see preceding section~\cite{Thouin2018,Neutzner2018}) yet distinct exciton states that are dressed differently by lattice vibrations, and this phonon dressing is yet again different from that experienced by photocarriers. 
\end{inparaenum}
We interpreted these phenomena as indicative of co-existing exciton polarons.

One immediate questions is: are the polaronic dressing phonons mere spectator modes or do they drive exciton dynamics? Emission from multiple excitons can be identified in the photoluminescence spectra with characteristic relaxation dynamics~\cite{straus2016direct, stranks2018influence}. Whilst the lowest lying state is dominant in the time-integrated spectrum, we observed population transfer dynamics from $X_B \rightarrow X_A$ in a few picoseconds~\cite{thouin2019polaron}. More importantly, the transfer dynamics are strongly thermally activated, with shorter transfer times at higher temperatures. We estimated the activation energy for the transfer to be $\sim 4$\,meV, which is also the energy of the dominant optical phonon mode identified in Ref.~\citenum{thouin2019phonon} (Fig.~\ref{fig_natmater}). This important observation suggests that the lattice modes are not mere spectators in the exciton relaxation dynamics, but are active in driving them. To rationalize this observation, we employed a mode-projection analysis~\cite{yang2014intramolecular}, which is a search algorithm that ranks the contributions from each of the experimentally observed Raman modes (with associated Huang-Rhys factors) to the nonadiabatic mixing and subsequently to the inter-exciton conversion. Our analysis suggested that the dominant contribution is from the 4-meV intra-lead-iodide-plane phonon mode, in agreement with the temperature-dependent inter-exciton transfer rate measurements. Importantly, we concluded that polaron-derssing phonons are active in driving $X_B \rightarrow X_A$ nonadiabatic relaxation~\cite{thouin2019polaron}.

\subsection{Multi-exciton correlations: exciton-exciton scattering and biexcitons}

In order to extract further insight into the consequences of distinct lattice dressing of $X_A$ and $X_B$ on their many-body interactions, we extended the 2D coherent measurements described above to a large range of exciton densities and temperatures in order to quantify the role of multi-exciton elastic scattering in optical dephasing~\cite{thouin2019enhanced}. Lineshape analysis of the zero-population-time rephasing coherent 2D excitation spectrum enables an unambiguous estimate of the homogenous linewidth~\cite{siemens2010resonance}, which is linked to the elastic scattering processes that lead to optical dephasing. We found that the temperature dependence of the dephasing rate of $X_A$ and $X_B$ is distinct, and reflects the role of different phonons in mediating exciton-exciton scattering. Perhaps more importantly, a simple Fr\"ohlich like scattering involving LO phonons may not be sufficient to rationalize the observed trend. We also found that the density dependence of the dephasing rate is measurably different for $X_A$ and $X_B$, and both dependences are approximately three orders of magnitude lower than in analogous measurements on single-layer transition metal dichalchogenides with comparable exciton and biexciton binding energies~\cite{moody2015intrinsic,martin2018encapsulation}. We interpreted this as a consequence of the polaronic nature of excitons in 2D-HIOPs: polaronic protection mitigates the effect of exciton-exciton elastic scattering on the dephasing rate in analogy to carrier scattering processes in electric transport~\cite{zhu2016screening,march2017four,zhu2015charge}.

\begin{figure}[ht]
	\centering
	\includegraphics[width=10cm]{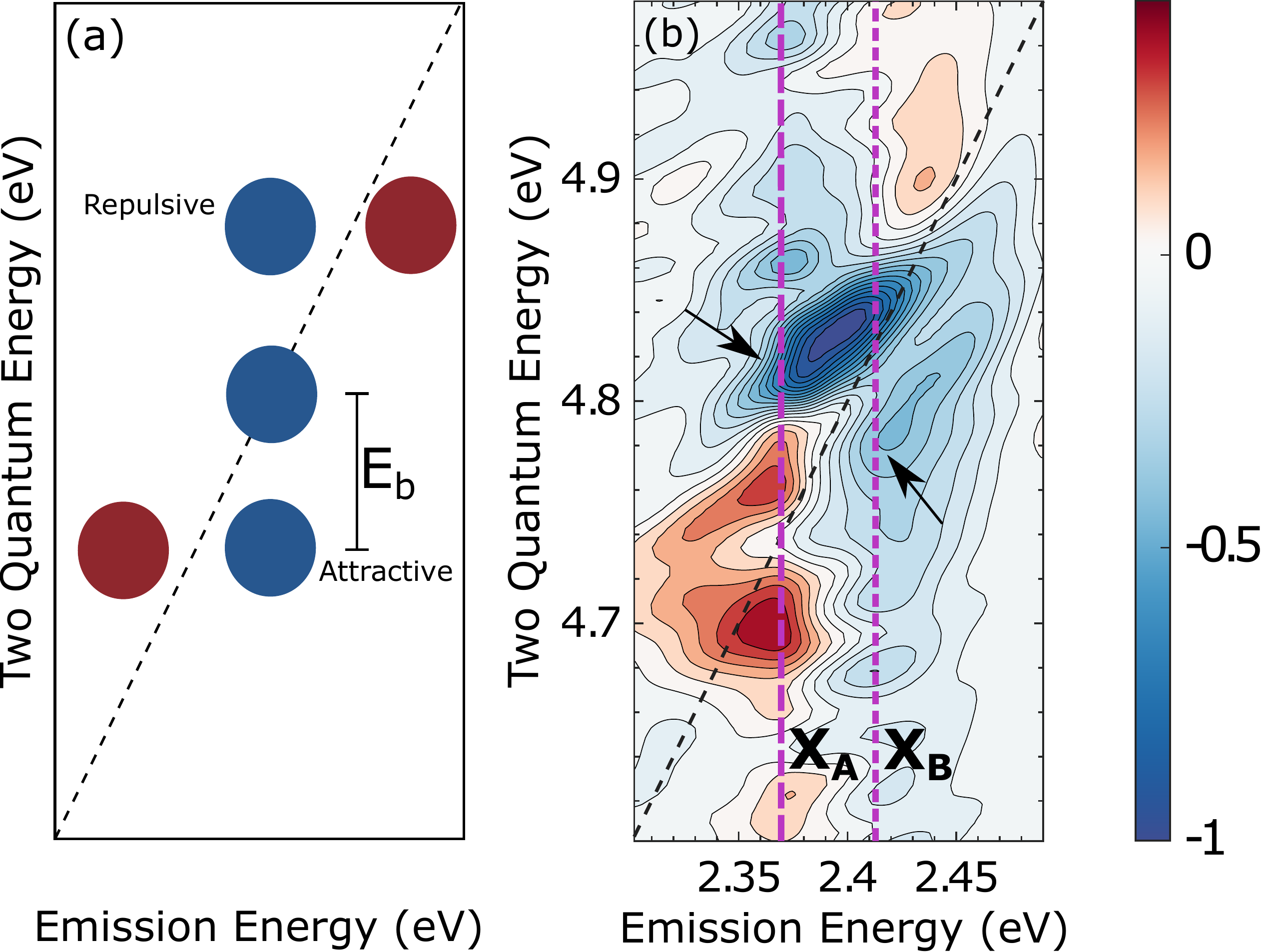}
	\caption{Real part of the two-quantum non-rephasing  2D coherent excitation spectrum of \ce{(PEA)2PbI4} at 5\,K for a one-quantum waiting time of 20\,fs. The dashed black diagonal line follows two-quantum energies at twice the emission energy. The vertical dashed lines display the peak energy of the two principal excitons in Fig.~\ref{fig:linear}. Figure extracted and modified from ref.~\citenum{Thouin2018}.}
	\label{fig4c}
\end{figure}

Protection from the elastic scattering process intriguingly does not hinder biexciton binding. By spectrally resolving two-quantum coherences in the 2D spectroscopic measurements described above, we reported clear biexciton signatures in 2D-HOIPs~\cite{Thouin2018}, as have others~\cite{elkins2017biexciton}. The intriguing observation was that the biexciton spectrum shows distinct binding interactions for $X_A$ and $X_B$. Shown in Fig.~\ref{fig4c}(a) is a representation of the expected two-quantum 2D coherent excitation spectrum for an exciton hosting both exciton-exciton binding and repulsive interactions. A diagonal feature, over the $y=2x$ two-quantum diagonal axis  is indicative of two excitons that experience no mutual interaction, leading to a two-quantum energy that is exactly twice the one-quantum energy. In the presence of attractive interactions (biexciton binding), a feature is observed below the diagonal along the two-quantum energy axis, shifted by the biexciton binding energy, while features observed above the diagonal reveal repulsive interactions (the energy of two excitons is higher than twice the energy of the single exciton). Shown in Fig.~\ref{fig4c}(b) is the two-quantum correlation spectrum, reproduced from Ref.~\citenum{Thouin2018}. The energies of $X_A$ and $X_B$ are indicated as dashed vertical lines. We observe a biexciton binding energy of $\sim 50$\,meV in the case of $X_B$, while it appears weaker for $X_A$, which is also subjected to substantial repulsive interactions revealed by resonances at two-quantum excitation energies of 4.86 and 4.95\,eV.  We once again interpret this within the optic of distinct lattice dressing of both excitons: we envisage that certain dressing phonons could promote exciton-exciton scattering, while others multiexciton binding. 

An apparent contradiction might be perceived by our interpretations put forth in this section. On the one hand, we have claimed polaronic protection to exciton-exciton elastic scattering. On the other hand, we claim that biexciton binding is strong, at least for $X_B$. We consider that this reflects the spatial extent of the polaronic distortion around excitons. If the inter-exciton separation is sufficient compared to the polaron radius associated with the electron-hole pair, dynamic screening by the lattice mitigates Coulomb-mediated elastic scattering. On the other hand, if the two-electron, two-hole spatial distribution is within the polaronic radius associated with that system, then biexciton binding interactions can be strong in this 2D system. This consideration would set constraints for the relative radii of of exciton polarons and their corresponding biexcitons and on the relevant spatial range of exciton-lattice coupling. 

In order to discuss the exciton-polaron hypothesis in further detail, we first summarize established formalisms for polarons in crystalline solids. We particularly invoke three relevant concepts in this discussion --- \begin{inparaenum}[(1)]
	\item the role of short-range lattice interactions and dimensionality in determining the polaron size, 
	\item the requirements for polaronic effects on excitons in the long-range interaction limit, and
	\item exciton self trapping in ionic lattices. 
\end{inparaenum}

\section{Polarons}

There are predominantly two kinds of polaronic effects in the context of semiconductors. The Fr\"ohlich formalism for large polarons is invoked commonly for carriers in polar and ionic semiconductors~\cite{frohlich1954electrons}, while the Holstein formalism for small polarons is used predominantly for molecular semiconductor crystals~\cite{holstein1959studies1,holstein1959studies2}. The primary distinction between the two cases stems from the spatial range of the Coulomb potential felt by carriers relative to the lattice parameters. In the case of Fr\"ohlich coupling, long-range interactions with lattice vibrations mediated by phonons modify the electronic structure in that the effective mass is increased while the Bloch-wave nature of the carrier is maintained. Further increase in the interaction strength can subsequenly lead to spatial localization of the carrier in the so-called small polaron limit. The essence of these two limits is quantitatively perceived via the polaron coupling constant $\alpha$ given in Eq.~\ref{Eq:alpha}, which can be estimated using measurable material characteristics such as dielectric permittivities at static and optical frequencies ($\epsilon_s$ and $\epsilon_\infty$), carrier effective masses ($m^*$), and the energy of the longitidinal optical phonon ($\hbar\omega_{LO}$) involved in the polaronic coupling~\cite{emin2013polarons, frohlich1954electrons}:  
\begin{equation}
    \alpha = \frac{e^2}{\hbar}\frac{1}{4\pi\epsilon_0} \sqrt{\frac{m^*}{2\hbar\omega_{LO}}}\left[ \frac{1}{\epsilon_\infty} - \frac{1}{\epsilon_s}\right].
    \label{Eq:alpha}
\end{equation}
Covalent solids where $\epsilon_s$ and $\epsilon_\infty$ are not vastly different tend to have an $\alpha < 0.5$ and host weak electron-phonon coupling, while materials with $\alpha > 3$, mostly due to polar lattice vibrations, host small-polaron-like excitations. Examples of the latter include metal-halides such as \ce{KCl} ($\alpha = 3.44$), \ce{CsI} ($\alpha = 3.67$) and perovskite structures such as \ce{SrTiO3} ($\alpha = 3.77$). In HOIPs, $\alpha$ can be estimated to be between 2 and 3 and thus the polaron coupling may be classified to be intermediate. 

While several experimental evidences exist that indirectly indicate the polaronic picture in these materials, there is still a need to develop direct probes that can unambiguously and quantitatively demonstrate polarons. Early theoretical works have demonstrated that optical absorption of polarons have characteristic lineshapes that can be analysed rigorously~\cite{devreese1972optical}. The presence of zero-phonon lines and the phonon replicas in the experimental infrared absorption spectra of materials, especially in organic semiconductors~\cite{voss1991substitution} and some superconductors~\cite{ruani1988dependence}, have enabled successful identification of small polarons. Magneto-absorption studies have revealed pertinent insights into the Fr\"ohlich coupling mechanisms~\cite{johnson1966polaron, devreese1989polaron}. There are several excellent reviews on these aspects~\cite{alexandrov2008polarons} and we do not intend to provide a exhaustive perspective on this topic here. We do note, however, that such characteristic signatures have not emerged yet in the case of metal-halide HOIP perovskites. 

Following Emin~\cite{emin2013polarons}, the total system energy of an electron in a deformable continuum under the adiabatic approximation can be written as
\begin{equation}
    E_{p} (L) = \frac{T_e}{L^2} - \frac{V_L}{L} - \frac{V_S}{L^D},
    \label{Eq_pol1}
\end{equation}
where $L$ is a dimensionless scaling factor that scales the position $\Vec{r}$ in the electronic wavefunction $\phi(\Vec{r})$ as $\Vec{r}/L$ and thus is related to the relative length-scale of the polaron. $T_e$ is the electronic kinetic energy, $V_L$ is the long-range interaction potential related to the Fr\"ohlich polaron coupling constant $\alpha$. The expression in Eq.~\ref{Eq_pol1} also considers the contributions from short-range interactions via the term $V_S$ and the effect on the dimensionality via the parameter $D$. In the presence of only long-range interactions, the minimum occurs at $L = 2T_e/V_L$ that describes the large polaron irrespective of the dimensionality. Inclusion of $V_S$ introduces the effect of dimensionality. With only short-range interactions, in the 3D case, one can obtain two minima at $L \rightarrow 0$ and at $L \rightarrow \infty$ that define the small-polaron and free-carrier limits, with an energetic barrier at $L = 3V_S/2T_e$ that is determined by the relative strengths of $V_S$ and $T_e$. In the presence of both short- and long-range interactions, the relative strength of $V_S$ and $V_L$ determines the polaron size. This is presumably the relevant regime for bulk HOIPs given the apparent range of $\alpha$ characterizing these materials.

In the case of 2D, the contribution from $V_S$ acquires similar functional dependence on $L$ as the kinetic energy term and thus the sign of $T_e - V_S$ determines the polaron size. Unlike the 3D case, in 2D, increase in the short-range interactions enables the formation of large polarons by reducing the kinetic energy contributions, before reaching the small polaron limit at $V_S > T_e$. Thus, we can infer that short-range interactions will primarily determine the electron-phonon coupling strengths in 2D lattices both in the large- and small-polaron limits.

\subsection{Exciton-phonon scattering problem}

Unlike uncorrelated electrons and holes, which carry a net charge, excitons are globally neutral quasi-particles. Thus, in principle, they may not necessarily be susceptible to the deformation potentials within the ionic lattices and may therefore be immune to polaronic effects. Exciton-phonon interactions can be represented as a total of the electron-phonon and hole-phonon interactions, represented as $\hat H_{QP-ph}$ in Eq.~\ref{Eq_phonon}, while the Coulomb interactions between electrons and holes is captured by $\hat H_{e-h}$ in Eq.~\ref{Eq_exc}. 
\begin{equation}
    \hat H_{e-h} = \frac{1}{N} \sum\limits_{pkk'} U (p,k,k') e_{p+k}^\dag h_{p-k}^\dag h_{p-k'}e_{p+k'}, 
    \label{Eq_exc}
\end{equation}
\begin{equation}
\hat H_{QP-ph} = \sum\limits_{k,q} \left( \gamma_e e_{k-q}^{\dag}e_k + \gamma_h h_{k-q}^\dag h_k \right) \left( b_q +b_{-q}\right), 
\label{Eq_phonon}
\end{equation}
where $\gamma_e$ and $\gamma_h$ are the coupling constants and are determined by the polaron coupling scenarios briefly described in the previous paragraphs and are related to the polaron coupling parameter ($\alpha$). 

For the sake of simplicity, let us initially consider a scenario where only long-range phonon interactions are present. Given that the nature of the interactions are equivalent to both valence and conduction-band states, one can deduce $\gamma_{e,q}$ = $-\gamma_{h,q}$ = $\gamma_q$ (note that this is unlike potentials created via deformations or acoustic lattice modes). The effective phonon interaction parameter that leads to an elastic scattering event within the 1s exciton band is then given by~\cite{ueta2012excitonic}
\begin{equation}
    \gamma_{1s\rightarrow 1s}(q) = \gamma_q\left(\left[1+(\xi_ea_Bq/2)^2 \right]^{-2} - \left[1+(\xi_ha_Bq/2)^2\right]^{-2} \right),
    \label{Eq_excph_scat}
\end{equation}
where $\xi_{e/h} = m_{e/h}/(m_e+m_h)$, $m_e$ and $m_h$ are electron and hole effective masses, respectively, and $a_B$ is the exciton Bohr radius. By energy and momentum conservation, a 1s exciton at $|\Vec{K}|\sim 0$, where $\Vec{K} = \Vec{k_e} + \Vec{k_h}$ is the exciton wavevector, can be scattered only via absorption of an optical phonon at $q \sim \sqrt{2(m_e+m_h)\omega_{LO}/\hbar}$. It is then evident from Eq.~\ref{Eq_excph_scat} that exciton-phonon coupling is significant only if $\xi_ea_Bq_0 \gg 1 \gg \xi_ha_Bq_0$, which translates as the criterion for Fr\"ohlich-like exciton-phonon scattering,
\begin{equation}
    \frac{m_h}{m_e} \gg \frac{E_B}{\hbar \omega_{L0}} \gg \frac{m_e}{m_h}. 
\end{equation}
Thus, materials with exciton binding energy $E_B \sim \hbar\omega_{LO}$ and with $m_e > m_h$ (or vice-versa) are subject to strong exciton-phonon scattering processes. Examples of such systems include silver-halides, thallous-halides and II-VI compounds~\cite{ueta2012excitonic}. A consequence of such scattering process is the appearances of vibronic lineshapes and phonon replicas in optical absoprtion and luminescence spectra. Alkali halides and cuprous halides, on the contrary, have large exciton binding energy, $E_B \gg \hbar \omega_{LO}$ and thus the 1s excitons are protected from LO phonon scattering processes. This can also be reformulated by invoking relative exciton and polaron sizes~\cite{ueta2012excitonic}. When the exciton size is large or comparable to the size of electron/hole-polarons, it will also be subjected to similar phonon interactions. On the contrary, if the exciton is much smaller than the polaron radius, it is unlikely to be scattered by the long range phonon interactions. Given the large exciton binding energies and equivalent electron and hole masses in 2D-HOIPs~\cite{Silver2018}, Fr\"ohlich-like interactions and thus polaronic effects should not be relevant for excitons. However, given that our recent works demonstrate that lattice coupling effects are indeed active and important~\cite{thouin2019phonon,thouin2019polaron,thouin2019enhanced}, we consider that one should move beyond the Fr\"ohlich picture to describe exciton polarons effectively in 2D-HOIPs. This highlights the need to consider an inter-play between short-range and long-range lattice coupling.

\subsection{Localization of excitons by phonon fields}

\begin{figure}[ht]
\centering
\includegraphics[width=\textwidth]{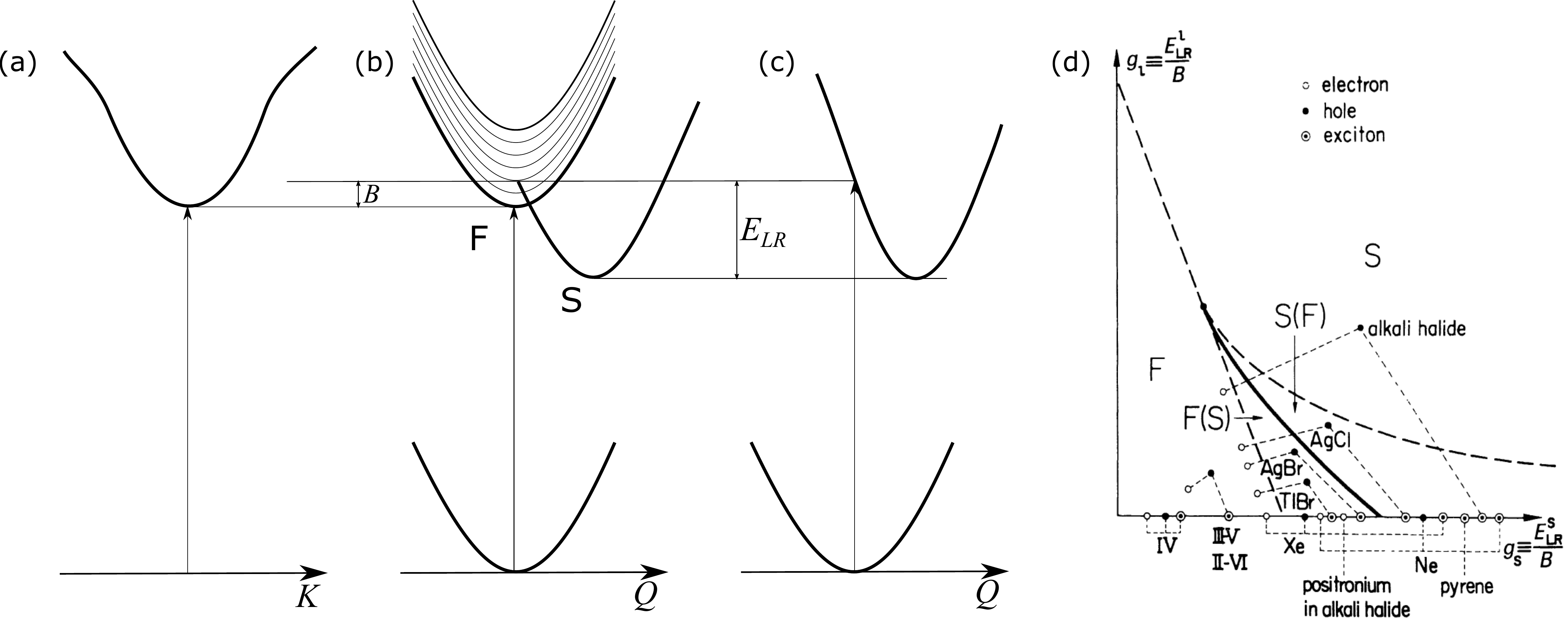}
\caption{Phase diagram showing stability of free (F) and self-trapped (S) carriers in the $g_l - g_s$, plane, where $g_l$ and $g_s$, are the long-range and short-range electron-phonon coupling constants defined in the text. Free and self-trapped carriers coexist in the F(S) and S(F) regions. The $(g_l,g_s)$ coordinates for electrons, holes, and excitons in several materials are noted. Reproduced with permission from ref.~\citenum{Toyozawa1981}}
\label{fig:phase2}
\end{figure}

Another perspective of the exciton-polaron problem has been proposed by Toyozawa~\cite{Toyozawa1981,ueta2012excitonic}, primarily within the strong electron-phonon coupling limit, and thus strictly valid for the Frenkel exciton case. In such a limit, the coupling constant may be defined as a ratio of the lattice relaxation energy ($E_{LR}$), that is the energy spent in re-organizing the lattice to accommodate the excitation, and the inter-site transfer energy ($B$). The energy of an exciton at the state $|\Vec{K}| = 0$ can now be represented along the lattice co-ordinates $\Vec{Q}$ as shown in Fig.~\ref{fig:phase2}(b). The potential energy surface (PES), and thus the equilibrated lattice configuration of the lattice-dressed state $S$ (self-trapped exciton), is shifted in the co-ordinate space with respect to the free exciton. Physical chemists will recognize the situation in Fig.~\ref{fig:phase2}(c) as analogous to that of molecular excitations, in which electron-vibrational coupling involving local nuclear modes modulates optical transition probabilities. The phonon coupling constant, an equivalent to the Huang-Rhys parameter, is written as $g = E_{LR}/B$. Photoexcitation can lead to lattice reorganization over relatively short lengthscales due to its localized nature via coupling to short-range lattice fluctuations, quantified by $g_s$ in Fig.~\ref{fig:phase2}(d). In a crystalline lattice, long-range optical phonons also couple to excitations and can be quantified by $g_l$. The coordinates $(g_s,g_l)$ define the phase space for generic carrier-lattice interactions, where we can identify a quantifiable range for free and self-trapped excitations. As shown in Fig.~\ref{fig:phase2}(d) and identified by Toyozawa~\cite{Toyozawa1981}, material systems can be marked over the phase diagram depending on the nature of lattice interactions. Conventional systems such as III-V and II-VI semiconductors have relatively weak lattice interactions, in the long and short range limits leading to \textit{free} excitations. Organic systems such as pyrene or rare gas crystals have substantially strong short-range interactions and thus host self-trapped excitations. Most of the metal-halides have substantial long-range lattice contributions (Fr\"ohlich-like) owing to their strongly ionic lattice, but they are also subjected to large short-range interactions due to the dynamic lattice fluctuations. This puts them at the conjunction of the free and self-trapped limits. 2D-HOIPs are no different in this context and a critical balance between the long- and short-range interactions plays a crucial role in determining the nature of photo-excitations. It is intriguing to note that octahedral distortions in 2D HOIPs, which enhance short-range interactions, have been correlated with broadband photo-luminescence characteristic of self-trapped states~\cite{cortecchia2016broadband} and large polaron binding energies~\cite{yin2017excitonic, cortecchia2017polaron, neukirch2018geometry}. 

\subsection{Excitons in 2D-HIOPs: The nature of polaronic coupling}

Based on this brief overview of some key concepts on polarons, in the context of 2D-HOIPs we can deduce that:
\begin{inparaenum}[(a)]
	\item short-range interactions will play an important role in determining the spatial extent of the polaronic wavefunction. Such short-range interactions may arise from the dynamic lattice fluctuations induced by the relative motion of the organic and inorganic sub-lattices~\cite{Thouin2018,dragomir2018lattice}, as well as from the disorder intrinsic to the ionic lattice. While phonons within the lattice plane may interact via the long-range Fr\"ohlich term, lattice motion across neighboring lattice planes will be perceived by the image charge as a short-range potential and thus contributes to the polaronic coupling.  
	\item Fr\"ohlich-like coupling may be irrelevant for excitons due to substantially large exciton binding energies in 2D-HOIPs. Given the clear experimental signatures of the polaronic effects on excitons, consideration of short-range interactions appears to be central to the exciton-polaron problem in 2D-HOIPs. 
	\item Exciton localization may be invoked in 2D-HOIPs. However, given that a model derived from a 2D Wannier picture successfully accounts for the optical absorption lineshape of 2D-HOIPs, with the caveat that the oscillator strength due to exciton absorption must be redistributed in multiple resonances with binding energy offset by multiples of $\Delta$~\cite{Neutzner2018}, we consider that the localization limit is unlikely to account quantitatively for exciton polarons in these materials because the assumptions of a Frenkel Hamiltonian are not satisfied. The hypothesis that we put forth in this perspective is that excitons in these materials are in an intermediate regime between the Fr\"ohlich and self-trapping limits. 
\end{inparaenum}
This poses substantial challenges for a rigorous, complete theoretical description of excitons in 2D-HIOPs, but our perspective is that this is a fundamentally important endeavor.

\section{The origin of the exciton spectral structure}

In Ref.~\citenum{Neutzner2018} we hypothesized that the spectral structure in Fig.~\ref{fig:linear} could reflect the importance of exciton polarons in 2D-HIOPs given that the difference in binding energy of multiple resonances is in the vicinity of the polaron binding energy, possibly reflecting distinct correlations of all possible binding combinations between electron- and hole-polarons and the unbound electrons and holes. Our report of distinct lattice dressing for $X_A$ and $X_B$ in Ref.~\citenum{thouin2019phonon} was important in establishing that polaronic effects are indeed reflected in exciton spectral structure. But can we rigorously speak of exciton polarons in 2D-HIOPs? In other words, are polaronic effects the origin of the observed exciton lineshape? We consider that further work outlined in the next section will be necessary to answer this question rigorously. In this section, we review the models put forth in the literature to rationalize the lineshape.

The first rationalization of the excitonic finestructure invoked large exchange interactions~\cite{Chen1988} in the 2D perovskite lattice, primarily enhanced by the dielectric confinement effects~\cite{Ema2006,Kitazawa2010}. Ema and coworkers have shown the existence of three non-degenerate excitonic states in 2D bromide-based hybrid perovskites~\cite{Ema2006,takagi2013}. These early experimental works were based on the analysis of low-temperature photoluminescence spectra, and they suggested that the excitonic manifold is composed of a low-energy dark state and two higher-lying bright states~\cite{Ema2006}. The molecular orbitals that contribute to the electronic bands are based within the \ce{PbBr6} octahedron. As discussed by Tanaka et al.~\cite{Tanaka2005}, the highest occupied orbitals (HOMO) and lowest unoccupied orbitals (LUMO) have $\Gamma_1^+$ and $\Gamma_4^-$ symmetries in the $O_h$ point group of the octahedron. The presence of crystal field effects, on top of spin-oribit interactions, splits the LUMO state into two bands with two $\Gamma_6^-$ symmetry and one band with $\Gamma_7^-$ symmetry. Based on a similar analysis, many early works suggested that the spin-oribit interactions that result in the LUMO splitting is thus responsible to the excitonic finestructure. These states have also been proposed to have their respective polarizations aligned parallel and perpendicular to the inorganic lattice planes. More intriguingly,  Ema et~al.~\cite{Ema2006} estimated a spin-exchange energy of approximately 28--32\,meV, similar to $\Delta$ in Fig.~\ref{fig:linear}.

We note, however, that we observe at least four equally spaced excitonic transition in the linear absorption spectra,~\cite{Neutzner2018} contrary to the three transitions in the PL spectra,~\cite{Kitazawa2010} which indicates that exchange interaction may not be the only viable origin for the finestructure. Moreover, Kataoka et~al.~\cite{Kataoka1993} did not observe any differences in the diamagnetic shifts of each of the excitons, further suggesting that spin exchange may not be the dominant origin of the finestructure. 

The excitonic finestructure may alternatively arise from a degeneracy-lifting mechanisms such as Rashba-Dresselhaus effects~\cite{manchon2015new}, which have been proposed in lead-iodide perovskites~\cite{stranks2018influence}. Large spin-orbit coupling due to the presence of lead and the absence of a center of inversion will result in lifting of spin degeneracies and splitting of the carrier bands~\cite{manchon2015new}. Todd et~al.~\cite{todd2018detection} have reported Rashba spin splitting energies that are 20 times larger than that of GaAs in 2D HOIPs. Consequently, excitonic states will also be split leading to a spectral finestructure. Zhai et~al.~\cite{zhai2017} have, in fact, reported a Rashba splitting of 40\,meV in \ce{(PEA)2PbI4} based on transient and quasi-steady-state absorption experiments, which appears to be consistent with our observation. However, we observed a lack of sensitivity of the finestructure to the thickness of the inorganic layer~\cite{Neutzner2018}, which determines the strength of the crystal field, and to the identity of the metal ion (similar spectral structure are observed in tin based systems), suggesting that Rashbha effects may not be origin of the finestructure. 

A few recent works have suggested that the observed spectral structure is a vibronic progression and not a finestructure composed of distinct excitonic states~\cite{straus2016direct, Straus2018a,giovanni2018coherent,mauck2019excitons}. Straus et al.~\cite{straus2016direct} reported transient dynamics similar to those reported by us~\cite{thouin2019polaron}, with a short-living emission band which was interpreted as non-Kasha emission from a vibrational manifold of a single exciton. While such an interpretation was certainly plausible given the data available, our observations in Refs.~\citenum{Thouin2018,thouin2019phonon,thouin2019enhanced} portrays that there is a more complex origin to the spectral lineshape with a non-negligible contribution from polaronic effects, and that the various resonances within the exciton lineshape have unique identity and do not arise from a single exciton.  

We underline that we find no reason to conclude that polaronic effects are the \emph{unique} contribution to the exciton lineshape, but do conclude that they are an important component of the physical phenomena that are manifested in the optical spectrum. Exchange interactions and Rashba-Dresselhaus effects may indeed co-exist with the type of lattice-coupling effects that we invoke in this perspective, and furthermore, lattice coupling effects may be a common element linking all of these phenomena. We have outlined above why we consider that each of these is not a unique mechanism that defines the spectral lineshape. Conversely, we have no reason to exclude their contribution. 

\section{Perspective}

Given the discussion of exciton polarons presented above, our perspective is that the intricate details of the fine structure in 2D-HOIPs can only be rigorously established via a complete theoretical treatment that would predict the full excitonic dispersion and that includes spin-orbit coupling effects~\cite{zhai2017}, exchange interactions~\cite{Ema2006,Kitazawa2010a, takagi2013influence}, many-body correlations~\cite{Thouin2018,thouin2019enhanced} and non-negligible yet complex polaronic effects~\cite{zheng1998,thouin2019phonon}. If our interpretation of the relevance of polaronic effects in excitonic properties holds, the interplay between long-range and short-range couplings puts polaronic effects in an intermediate regime towards a localization limit, and microscopic detail will be crucial in this development. This is a non-trivial challenge in condensed-matter theory. Recently, Sio et~al.\ have developed a rigorous \textit{Ab initio} framework to study polarons in semiconductors~\cite{Sio:2019aa,Sio:2019ab}. That work uses density-functional perturbation theory to solve a secular equation involving phonons and electron-phonon coupling. This approach is analogous to that invoking the Bethe-Salpeter equation for excitons in the absence of electron-phonon coupling, and we therefore consider that it presents unprecedented opportunities for the most rigorous examination of the nature of exciton polarons in 2D-HOIPs. Nevertheless, we recognize that this approach is far from straightforward. 

From an experimental perspective, we consider that there is substantial scope to implement ultrafast structural probes in conjunction with electronic spectroscopies to examine in detail exciton-polaron dynamics. In particular, ultrafast electron diffraction and scattering techniques are rapidly developing towards implementable tools for materials science~\cite{carbone2010real,konstantinova2018nonequilibrium}. Specifically, the dynamics of electron phonon coupling can now be mapped in crystalline systems~\cite{stern2018mapping}, and we put forth that these techniques would provide unprecedented microscopic detail. We consider that there is substantial scope for advanced time-resolved vibrational spectroscopies performed in conjunction with electronic spectroscopies~\cite{munson2019lattice}. 

Rigorous understanding of exciton polarons in 2D-HOIPs is fundamentally important not only in the context of the development of their semiconductor physics, but generally for a much broader class of materials in which multiparticle coupling and many-body effects are mediated by interactions with a highly dynamic lattice. In conventional superconductors, for example, Cooper pairs result from lattice-mediated binding of electrons that would otherwise experience net repulsion~\cite{koschorreck2012attractive}. Another example of perhaps greater contemporary significance points to non-conventional quantum materials (many of them with perovskite crystal structures), in which, for instance, the lattice plays an important role in mediating correlated quantum phenomena involving spin-orbital entanglement when spin-orbit coupling is strong~\cite{Witczak-Krempa:2014aa}. We consider that 2D-HOIPs are an ideal test bed for developments in condensed-matter theories that seek a rigorous description of multi-particle correlations, including all of the elements that are also important in quantum materials --- polaronic effects and related Jahn-Teller-type lattice distortions, spin-orbit coupling, and multi-particle correlations. HOIPs permit a clear window into this broadly important materials science via their very clear optical properties, providing a simple experimental access to a multidimensional materials parameters space involving static and dynamic structure, dimensionality, chemical composition, and spin-orbit coupling strength, for example. Beyond a perceived relevance as semiconductors for optoelectronics, HOIPs are a fundamentally valuable model class of materials for substantial advances in the undersstanding of many-body effects that are critical in condensed-matter and materials physics, chemical physics, and materials chemistry.

\section{Author Information}

\subsection{Corresponding Authors}
$^*$E-mail: srimatar@wfu.edu (A.R.S.K.) \\
$^*$E-mail: carlos.silva@gatech.edu (C.S.)

\subsection{Notes}
The authors declare no competing financial interest.

\subsection{Biographies}
\textbf{Ajay Ram Srimath Kandada} received a Ph.D.\ in Physics from Politecnico di Milano, Italy in 2013 and currently he is a Marie Sklodowska Curie fellow at the Italian Institute of Technology. Previously, he was a post-doctoral scholar at University of Montreal, Canada and Georgia Institute of Technology, USA. Starting January 2020, he will be an Assistant Professor in Physics at Wake Forest University, USA. His research interests include advanced optical spectroscopy of semiconductors and photo-excitation dynamics in hybrid lead-halide perovskites.

\noindent
\textbf{Carlos Silva} earned a Ph.D.\ in Chemical Physics from the the University of Minnesota in 1998, and was then Postdoctoral Associate in the Cavendish Laboratory, University of Cambridge. In 2001 he became EPSRC Advanced Research Fellow in the Cavendish Laboratory, and Research Fellow in Darwin College, Cambridge. In 2005, he joined the Universit\'e de Montr\'eal as Assistant Professor, where he held the Canada Research Chair in Organic Semiconductor Materials from 2005 to 2015, and a Universit\'e de Montr\'eal Research Chair from 2014 to 2017. He joined Georgia Tech in 2017, where he is currently Professor with joint appointment in the School of Chemistry and Biochemistry and the School of Physics, and Adjunct Professor in the School of Materials Science and Engineering. His group focuses on optical and electronic properties of organic and hybrid semiconductor materials, mainly probed by nonlinear ultrafast spectroscopies and quantum-optical methods.

\begin{acknowledgement}
We are indebted to all of our collaborators and co-authors in the development of this work, but primarily F\'elix Thouin, Daniele Cortecchia, Stefanie Neutzner, Annamaria Petrozza, Claudio Quarti, David Beljonne, David Valverde Ch\'avez, Ilaria Bargigia, and Eric Bittner.  A.R.S.K. acknowledges funding from EU Horizon 2020 via Marie Sklodowska Curie Fellowship (Global) (Project No.\ 705874). C.S. acknowledges funding from the National Science Foundation, Directorate for Mathematical and Physical Sciences, Division of Materials Research (Award number 1904293 and 1838276), and for support from the School of Chemistry and Biochemistry and the College of Science of Georgia Institute of Technology. 
\end{acknowledgement}





\providecommand{\latin}[1]{#1}
\makeatletter
\providecommand{\doi}
{\begingroup\let\do\@makeother\dospecials
	\catcode`\{=1 \catcode`\}=2 \doi@aux}
\providecommand{\doi@aux}[1]{\endgroup\texttt{#1}}
\makeatother
\providecommand*\mcitethebibliography{\thebibliography}
\csname @ifundefined\endcsname{endmcitethebibliography}
{\let\endmcitethebibliography\endthebibliography}{}

\end{document}